\documentclass{article}
\usepackage[utf8]{inputenc}
\usepackage{graphicx}
\usepackage{multicol}
\usepackage{caption}
\usepackage{amsmath}
\usepackage{hyperref}
\usepackage[a4paper, left=0.75in, right=0.75in, top=0.8in, bottom=0.8in]{geometry}
\usepackage{float}
\usepackage{authblk}

\title{Parameter Analysis and Optimization of Layer Fidelity for Quantum Processor  Benchmarking at Scale}
\author{Maria Jose Lozano Palacio\thanks{mlozano@ibm.com}}
\author{Hasan Nayfeh}
\author{Matthew Ware}
\author{David C. McKay}
\affil{\small\textit{IBM Quantum, IBM T.J. Watson Research Center, Yorktown Heights, New York 10598, USA}}

\begin{document}
\maketitle

\begin{abstract}
With the continued scaling of quantum processors, holistic benchmarks are essential for extensively evaluating device performance. Layer fidelity is a benchmark well-suited to assessing processor performance at scale. Key advantages of this benchmark include its natural alignment with randomized benchmarking (RB) procedures, crosstalk awareness, fast measurements over large numbers of qubits, high signal-to-noise ratio, and fine-grained information \cite{lf}. In this work, we extend the analysis of the original layer fidelity manuscript to optimize parameters of the benchmark and extract deeper insights of its application. We present a robust protocol for identifying optimal qubit chains of arbitrary length $N$, demonstrating that our method yields error per layered gate (EPLG) values $40 \% - 70 \%$ lower than randomly selected chains for $N=100$ qubits. We further establish layer fidelity as an effective performance monitoring tool, capturing both edge-localized and device-wide degradation by tracking optimal chains of length 50 and 100, and fixed chains of length 100. Additionally, we refine error analysis by proposing parameter bounds on the number of randomizations and Clifford lengths used in direct RB fits, minimizing fit uncertainties. Finally, we use layer fidelity to analyze the impact of varying gate durations on layered two-qubit (2Q) errors, showing that prolonged gate times leading to idling times significantly increase these quantities. These findings extend the applicability of the layer fidelity benchmark and provide practical guidelines for optimizing quantum processor evaluations.
\end{abstract}

\begin{multicols}{2}

\section{Introduction}
\label{sec:introduction}

The development of quantum benchmarks allows for the evaluation and comparison of system performance across different devices and technologies. While average device performance is often inferred from randomized benchmarking (RB) protocols, these unstructured approaches do not necessarily reflect the layered or algorithmic nature of many near-term quantum workloads. Many near-term algorithms, such as the variational quantum eigensolver (VQE), the quantum approximate optimization algorithm (QAOA), and Trotterized dynamics rely on the repeated execution of layered gate operations (see \cite{review} for a review). Layered structures are also advantageous for other approaches: they connect naturally to Pauli learning and to error mitigation strategies that rely on rerunning circuits with different noise profiles \cite{error_cancellation, noisy}. Together, these connections provide compelling motivation for benchmarks that evaluate performance in a layered framework.

Several benchmarking methods have incorporated structured circuits to assess processor performance, including quantum volume (QV) \cite{qv}, cross-entropy benchmarking (XEB) \cite{xeb}, mirror RB \cite{mirrorrb}, inverse-free (binary) RB \cite{inversefreerb}, and cliffordization \cite{seth}. Random-circuit protocols such as QV and XEB resemble realistic workloads and provide system-level measures of performance. However, they rely on classical simulation to compute expected outcomes, which limits scalability as the number of qubits increases and often requires extensive sampling to obtain reliable statistics. In addition, QV is typically reported as a pass/fail test rather than a continuous metric, making it less sensitive to incremental improvements in device performance. Clifford-based approaches such as mirror RB and inverse-free RB avoid this simulation overhead and efficiently estimate aggregate quantities such as error per layer or per circuit block. While useful for characterizing overall processor quality, these methods still report global metrics and provide limited insight into which specific gates or regions of the device dominate the observed errors.

For superconducting processors, where error rates can vary significantly across the chip, coarse or global metrics can hide important differences between regions of the device. A useful holistic benchmark should therefore scale to large numbers of qubits while still revealing which specific gates or areas are responsible for the observed errors.

Layer fidelity \cite{lf} directly addresses this need by combining scalability with fine-grained diagnostics. It measures the fidelity of a connected set of two-qubit (2Q) gates over $N$ qubits by extracting gate errors using simultaneous direct RB \cite{simulRB,directrb} in disjoint layers. This approach aligns naturally with standard RB \cite{rb1,rb2} procedures, captures crosstalk through parallel execution, enables rapid measurements across many qubits, and maintains a high signal-to-noise ratio, while still preserving access to individual gate-level information. Given these advantages, further analysis is valuable to extract deeper insights into its applications. In this work, we extend the initial analysis in \cite{lf} by developing methods for identifying optimal qubit chains, establishing a stability monitoring tool, refining benchmark parameters, and analyzing the impact of varying 2Q gate durations.

\section{Using Layer Fidelity in practice}
\label{sec:motivation}

In addition to reporting on the holistic performance of a device, layer fidelity can also be a practical way to find the most performant regions of a processor. By measuring simultaneous direct RB across connected sets of qubits, the protocol allows us to compare different parts of the chip under realistic workload considerations, such as parallel execution and timing constraints \mbox{\cite{lf}}, and identify the regions with the highest performance. 

This is especially important for superconducting processors, where error rates vary across the device. As a result, performance depends strongly on the chosen qubit layout, making it important to identify the optimal regions of the processor. These high-performing regions are not only preferred targets for running experiments, but also provide a natural reference for monitoring stability over time and complement existing system-level benchmarks.

In this work, we focus on one-dimensional qubit chains as a simple and practical unit for evaluating the performance and stability of device regions. Many relevant workloads—such as simulations of one-dimensional Heisenberg \mbox{\cite{1d_p2}}, transverse-field Ising \mbox{\cite{1d_p1}}, and Hubbard models \mbox{\cite{1d_p3}}—naturally map to this geometry, and the approach we propose can be extended to more general connectivity patterns. In addition, chains provide several practical advantages that make them well suited for the benchmarking protocol. While layer fidelity can be evaluated on any connected set of qubits, chains simplify the partitioning of gates into disjoint layers, capturing crosstalk effects efficiently \mbox{\cite{lf}}. This is because a chain requires only two alternating layers—one starting at $Q0$ (even sites) and one at $Q1$ (odd sites). Furthermore, data on longer chains can be easily re-analyzed to provide the layer fidelity of any subchain.  

Therefore, using chains as the main case study, we use layer fidelity both to identify optimal device regions and to monitor their stability. The next two sections describe our protocol for locating these chains efficiently and for tracking device performance across time. While we demonstrate these methods on a superconducting architecture, the underlying concepts are hardware agnostic.

\section{Protocol for finding the optimal 1D chain of N qubits}
\label{sec:protocol}

In this work, we target chains of $N=100$ qubits. This length covers a large fraction of the processors considered here—127, 133, and 156 qubits—and is representative of the scale relevant to near-term utility applications. At the same time, the number of possible connected subsets grows rapidly with $N$, making it infeasible to physically run layer fidelity on all candidates. For example, a 156-qubit heavy-hex architecture (Heron R2) has 35,804,119 distinct chains of this length. Due to this combinatorial complexity, heuristic methods are required to efficiently estimate optimal chains over all possible subsets. The approach we propose below generalizes to other values of $N$, provided that connected chains of the desired length exist.

Our approach for identifying optimal chains uses single-qubit (1Q) and two-qubit (2Q) gate errors to prescreen the device and locate high-performing regions. We consider two strategies for measuring these gate errors that feed into our prescreening algorithm -- isolated and grid. In both cases, characterization experiments are performed in advance and the resulting metrics guide the subsequent search for the best $N$-qubit chain.

The isolated strategy measures gate errors from isolated RB, a variant of simultaneous RB \cite{simulRB} where all 2Q gate pairs are separated by at least two idle qubits and no barriers are enforced. In the worst case, obtaining all isolated gate errors would require $O(N)$ experiments, but this number is lower in practice given the batching specified above. This setup naturally reduces some forms of crosstalk, such as ZZ interactions between coupled qubits in fixed architectures. Since isolated RB does not apply barriers, it also does not enforce the constraint of running every RB sequence at the maximal gate duration  \cite{lf}. To account for this, we apply a small penalty to gate pairs with longer durations in the cost function defined in formula (1). 

The grid strategy, in contrast, derives gate errors from the layer fidelity RB protocol run on a predefined grid of qubit chains (see Fig.~\ref{fig:fig1} (a) and this \href{https://github.com/qiskit-community/qiskit-device-benchmarking/blob/main/notebooks/layer_fidelity_placement.ipynb}{notebook} for an example). The grid consists of two sets of vertical and horizontal chains, ensuring that all 2Q gate pairs on the device are included. When a 2Q gate pair appears in both configurations, we take the average of the two gate errors. In a device with $J$ couplers per qubit, these experiments can be grouped into $O(J)$ batches, making the strategy scalable to larger devices. The layer fidelity RB protocol enforces the constraint that a timing barrier must be applied to all gates in a layer, leading to potential idle times for shorter gates. As such, a limitation of the grid strategy is that it can introduce length-dependent effects in a way that is not always representative of the final optimal 100-qubit chain. For example, the grid strategy will always run some grid chains at the maximal gate duration of the device, but there is no guarantee that the longest gate will be part of the optimal 100-qubit chain, so the length-dependent

\begin{center}
  \centering
  \includegraphics[width=\linewidth]{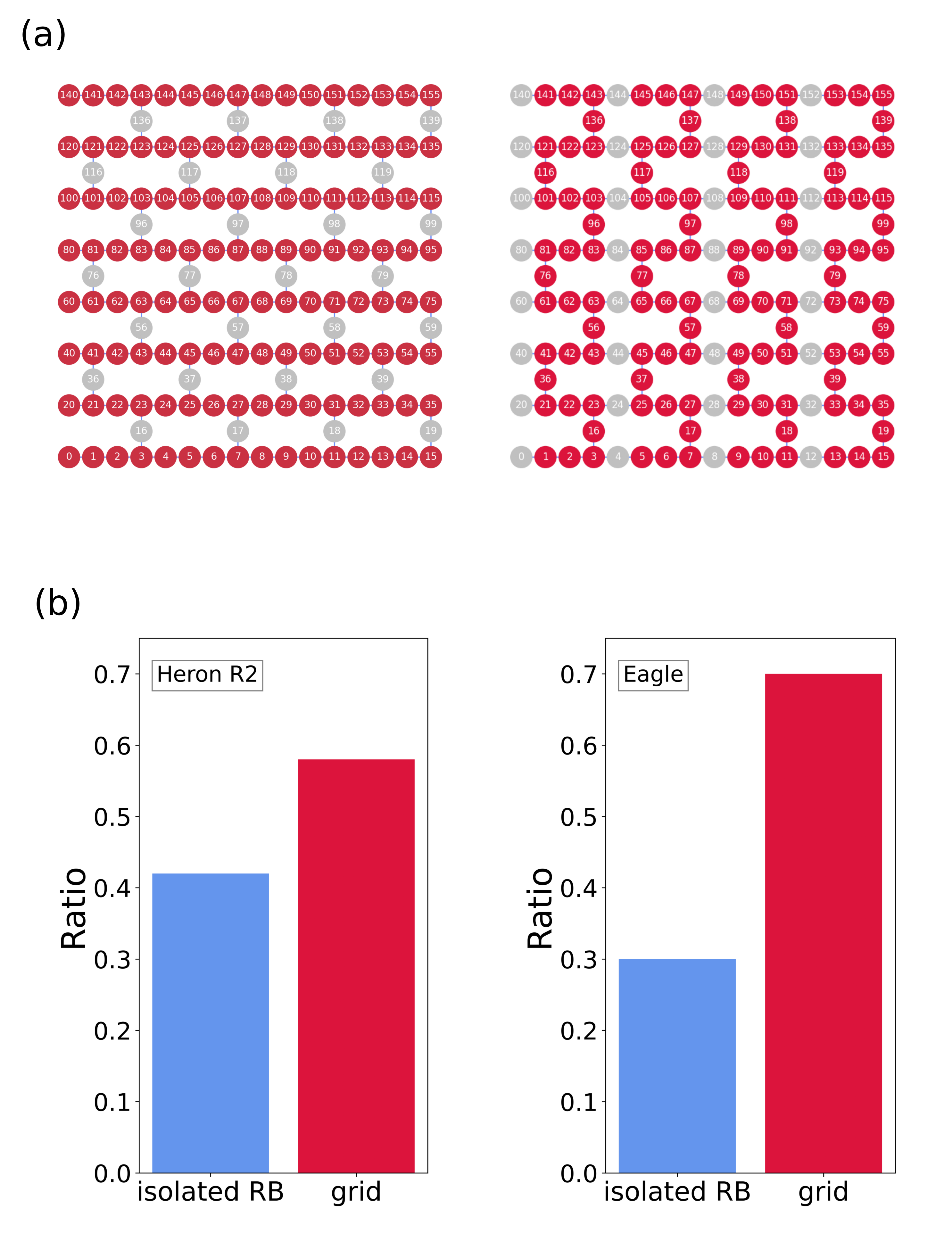}
    \captionsetup{type=figure}
    \captionof{figure}{ (a) Horizontal and vertical grid configurations for a Heron R2 processor. (b) Frequency of occurrence of the optimal $N=100$ qubit chain: isolated RB strategy (blue) vs grid strategy (red) for a Heron R2 ($ibm\_marrakesh$) and Eagle R3 ($ibm\_cusco$). The grid strategy shows a higher frequency of occurrence in both processors, highlighting the importance of considering crosstalk in the evaluation of the optimal device regions. Note: In this study, the backend $ibm\_cusco$ was chosen as opposed to other Eagle R3 backends because its 2Q gates had an equal gate duration across the entire chip, so one could isolate crosstalk effects from gate duration effects. }
  \label{fig:fig1}
\end{center}

effects this long gate introduces might lead to an increase of the effective 2Q error of other gates (see Section \ref{sec:duration} for a study on this effect). 

Once the 1Q and 2Q gate errors are obtained, both the isolated and grid strategies follow the same procedure to identify the highest-performing chain as visualized in Fig.~\ref{fig:fig2}. 

\begin{enumerate}
    \item We define a cost function $LF_{cost}$ as the product of individual process fidelities
    $F_j$ along some chain of size $N$
    
    \begin{equation}
    LF_{cost} = \prod_{j} F_j
    \end{equation}

    where $F_j = \frac{(1 - \epsilon_j)(d+1) - 1}{d}$ for all gate errors $\epsilon_j$ on the chain in question (this includes all gates across disjoint layers). Here, $d$ is the dimension of the decay space, where $d=2$ for 1Q errors (located at both ends of the chain) and $d=4$ for 2Q errors (located within the chain).

    The $N$-qubit chain with the highest cost function value among all possible candidates is designated as chain $A$. Note that finding the cost function of all chains on a device can be quite memory intensive, so we rely on a depth-first-search algorithm to find the chains with the highest cost function.
    \item To allow flexibility in chain selection, we identify additional candidates by selecting the next $x$ chains with the highest cost function values. Here, $x$ is an arbitrary number (defaulted to 15). This is important because, while chain A has the highest predicted performance, other chains with slightly lower cost function values may yield higher layer fidelity values when experimentally measured. 
    \item From this set of $x$ chains, we choose two chains, $B$ and $C$, that have the fewest overlapping qubits with chain A and have fairly trivial differences among them. Specifically, chain $B$ is the chain that has the largest number of non-overlapping qubits relative to chain $A$, while chain $C$ has the second largest. This produces a total of three chains.
    \item Since we are using two strategies, the isolated strategy and the grid strategy, we must apply this procedure twice. That is, we must find the best three estimated chains for each of the two methods. This leads to a final set $S=[A,B,C,D,E,F]$, where $A$, $B$, and $C$ come from the grid strategy, and $D$, $E$, and $F$ come from the isolated strategy. Here, chain $A$ represents the optimal chain identified by the grid strategy, while chain $D$ is the optimal chain identified by the isolated strategy.
    \item We then run the layer fidelity protocol on all chains in set $S$ and compare their performance. The best $N$-qubit chain for the device is defined as the chain within set $S$ that exhibits the lowest average error per layered gate value, EPLG, which can be directly determined from the layer fidelity value, LF, of the full layer. 
    
    The initial layer fidelity manuscript \mbox{\cite{lf}} defines the layer fidelity of a disjoint layer as the product of all process fidelities in that layer

    \begin{equation}
    LF_m = \prod_{j} F_{j,m},
    \end{equation}

    where $F_{j,m} = \frac{1 + (d^2 - 1)\alpha_j}{d^2}$ is the process fidelity of gate $j$ in the $m$th disjoint layer, $\alpha_j$ is the RB decay rate, and $d=2$ and 1Q gates or $d=4$ for 2Q gates. The full layer fidelity is

    \begin{equation}
    LF = \prod_{m} LF_m,
    \end{equation}
    
    \begin{center}
    \centering
      \includegraphics[width=\linewidth]{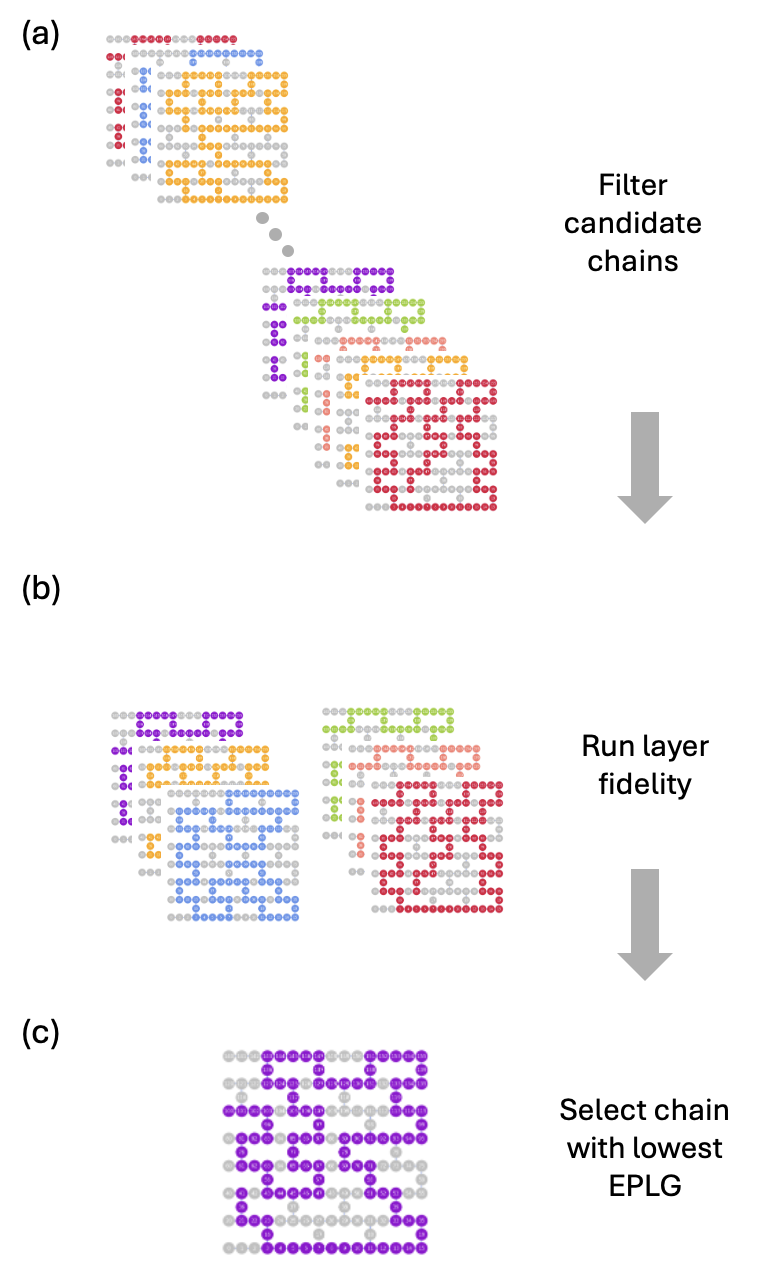}
      \captionsetup{type=figure}
      \captionof{figure}{Diagram of the protocol to find the optimal $N=100$ qubit chain. (a) For each strategy, isolated RB and grid, the top x chains are filtered from all candidate chains according to the cost function $LF_{cost}$ (x is defaulted to 15). (b) For each strategy, we select the top 3 chains from the subset of x chains. Chain 1 is the one with the highest $LF_{cost}$ and chains 2 and 3 have the least overlapping qubits with chain 1. (c) The layer fidelity is measured for each of the chains found on the previous step and the chain with the lowest EPLG is reported as the optimal chain.}
      \label{fig:fig2}
    \end{center}

    which can be used to calculate EPLG -- a normalized quantity with respect to the total number $n_{2q}$ of 2Q gates. We define EPLG to be an average 2Q gate error and so it has a prefactor of $\frac{d}{d+1}=\frac{4}{5}$.

    \begin{equation}
    EPLG = \frac{4}{5}(1 - LF^{1/n_{2q}})
    \end{equation}

\end{enumerate}

A key test of our protocol is whether it outperforms a randomly selected chain. To evaluate this, we compare the best $N=100$ qubit chain identified by our method to three randomly selected $N$ qubit chains on two processors: Heron R2 ($ibm\_marrakesh$, 156 qubits) and Eagle R3 ($ibm\_brisbane$, 127 qubits). The results 

\begin{center}
  \centering
  \includegraphics[width=\linewidth]{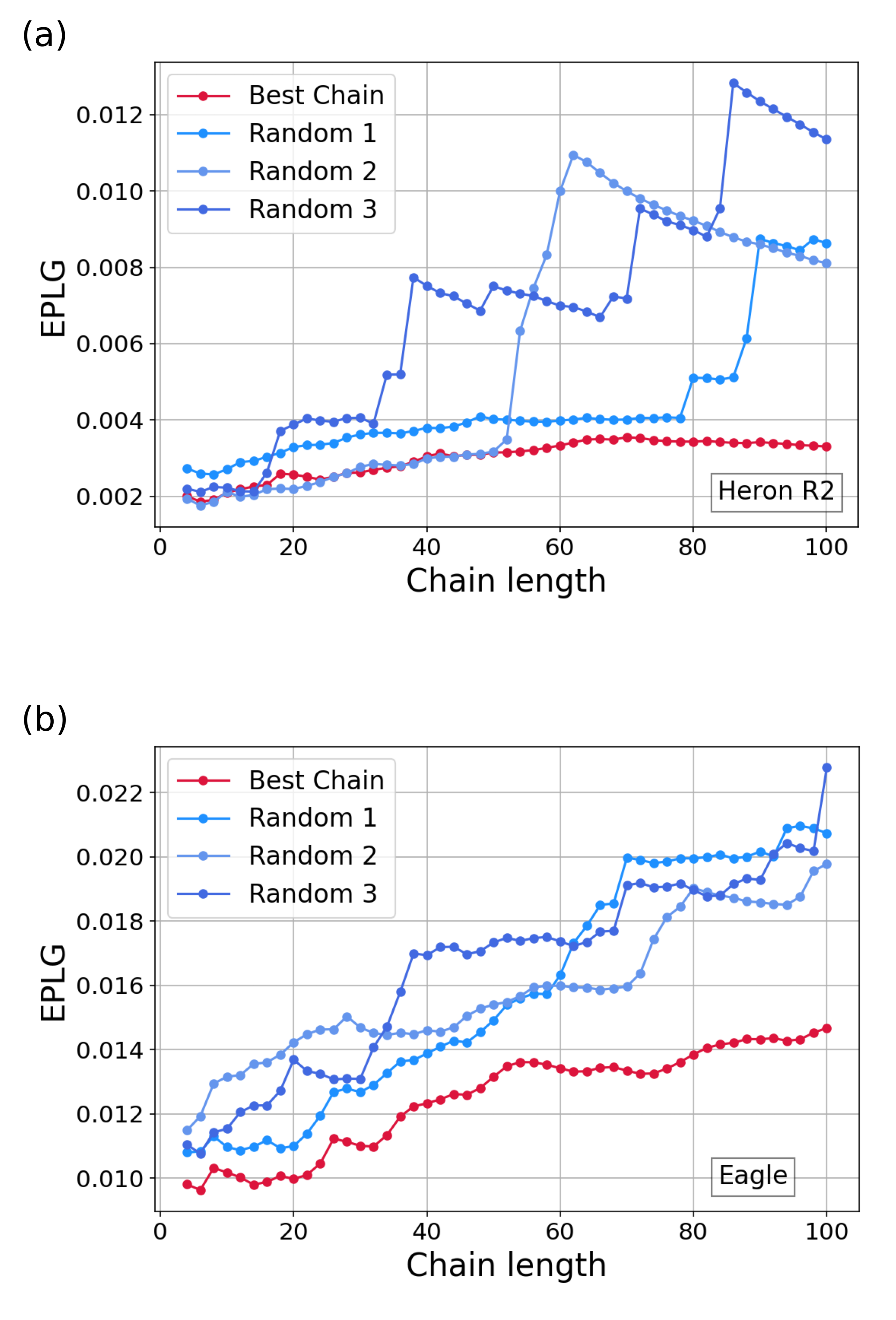}
  \captionsetup{type=figure}
  \captionof{figure}{ EPLG vs chain length for the optimal $N=100$ qubit chain (red) and 3 random $N$ qubit chains (blue) on (a) Heron R2 ($ibm\_marrakesh$)  and (b) Eagle R3 ($ibm\_brisbane$). The optimal chain shows superior performance in both cases.}
  \label{fig:fig3}
\end{center}

demonstrate a clear advantage of the proposed method, as the estimated best chains significantly outperformed their random counterparts. Specifically, the best chains identified by our protocol measured EPLG values 70\% and 40\% lower than the best performing random chains for Heron R2 and Eagle R3, respectively (see Fig.~\ref{fig:fig3} (a) and (b)).

In addition to comparing against random chains, we also examine the relative performance of the grid and isolated strategies. We count the frequencies of occurrence for each chain and strategy and normalize them by the total counts. By construction, one might expect chain $A$, selected by the grid strategy, to consistently yield the lowest EPLG within set $S$. However, a detailed comparison of the EPLG of chain $A$ against those of chains $B, C, D, E,$ and $F$ reveals that the highest-performing chain can originate from other candidates in the set as well. Daily data collected over 104 days shows that on Eagle R3 processors, the grid strategy identified the best-performing chain in $70\%$ of cases, with chain $A$ exhibiting the lowest EPLG in most instances. On Heron R2 processors, the performance of both strategies was more evenly distributed: the grid strategy identified the best-performing chain in $60\%$ of cases, with chain $A$ exhibiting the lowest EPLG in a third of those cases (see Fig.~\ref{fig:fig1} (d)). These trends are consistent with device characteristics: the grid strategy provides more accurate predictions when crosstalk is significant (Eagle R3), while both strategies perform similarly when crosstalk is reduced (Heron R2) \cite{heronr2.2}.

Taken together, these observations suggest that the grid strategy should be the default choice in most settings. Because it evaluates gates under simultaneous execution, it more closely reflects the conditions encountered in layered circuits and therefore provides a more reliable predictor of chain performance. The isolated strategy remains useful when rapid characterization is desired or when crosstalk is negligible, since it relies only on standard isolated 1Q and 2Q RB metrics that are often already available (e.g., through backend calibration data on Qiskit) and does not require additional experiments. However, when predictive accuracy is the primary objective, we recommend using the grid-based projections to select candidate chains.

While these projections identify high-performing chains at a single time point, reliable device operation requires evaluating their consistency over time. Accordingly, we monitor whether chain performance remains stable over days to months of operation.

\section{Using EPLG to track device stability over time}
\label{sec:monitoring}

To quantify this stability, we track EPLG as a function of time, which allows us to detect fluctuations in system performance and identify potential degradations. We consider two complementary cases. The first analyzes EPLG for the optimal chain identified at the time of each layer fidelity run, providing insight into how the best-performing regions of a device evolve. The second focuses on EPLG for a fixed chain, highlighting the long-term stability of a specific region.

In the first case, we run layer fidelity once per day over a 100-day period, recording EPLG values for the best $N=100$ and $M=50$ qubit-long chains on a device. The choice of these lengths is motivated by both practical and diagnostic considerations. A 100-qubit chain spans a large fraction of current superconducting processors and reflects the scale relevant for utility-scale experiments, where workloads typically occupy tens to hundreds of qubits. The shorter 50-qubit chain serves as a complementary probe that samples a smaller but still substantial portion of the device. Because the $M$-qubit chain contains fewer couplers, it can typically avoid isolated low-performing gate pairs; its EPLG therefore reflects changes that affect many qubits simultaneously. The longer $N$-qubit chain, by contrast, spans many more couplers and is consequently more sensitive to individual underperforming edges.

As a result, a simultaneous increase in both chain lengths indicates system-wide degradation, whereas an 

\begin{center}
  \centering
  \includegraphics[scale=0.85]{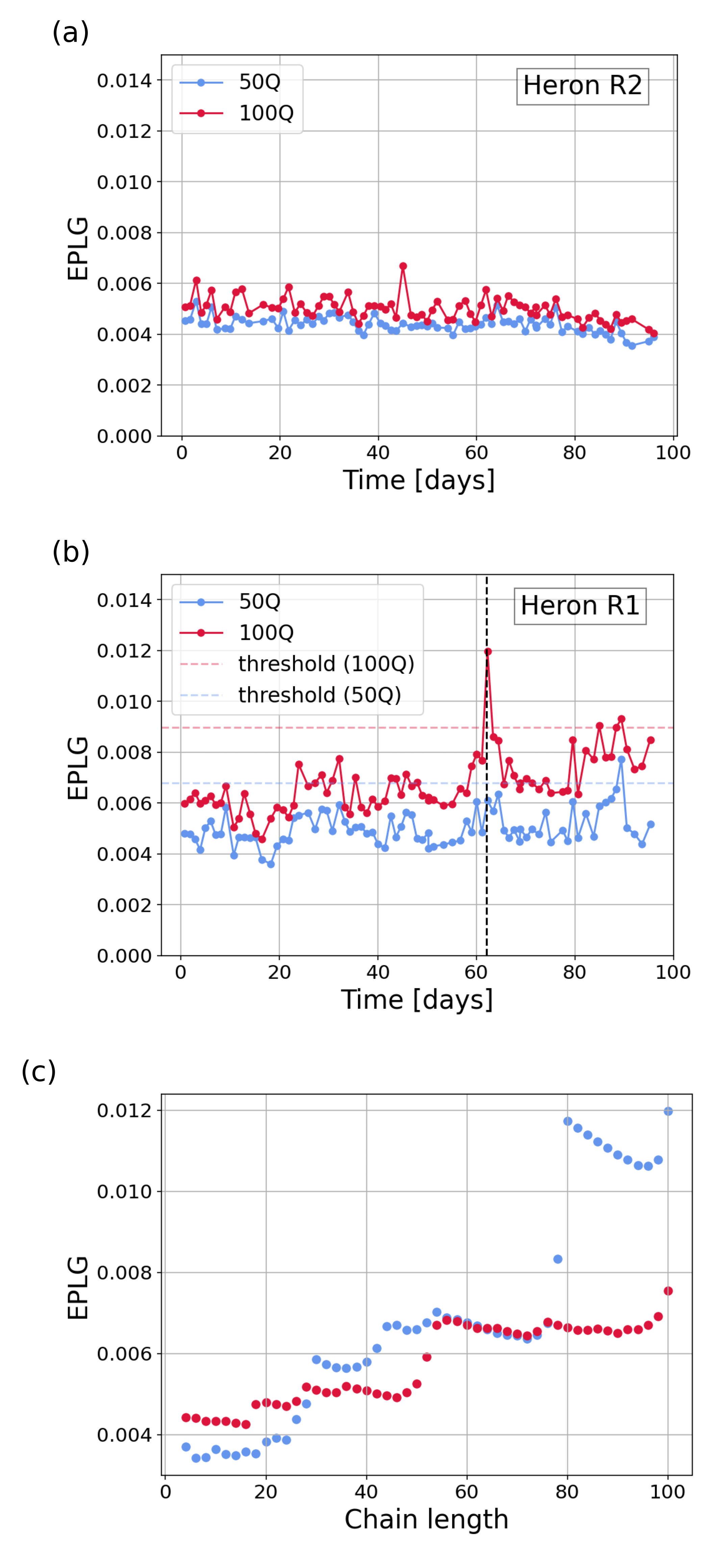}
  \captionsetup{type=figure}
  \captionof{figure}{EPLG vs time for (a) Heron R2 ($ibm\_fez$) and (b) Heron R1 ($ibm\_torino$) for chains of length 50 (blue) and 100 (red). On day 62, the EPLG of the 100-qubit chain on Heron R1 increases to $1.2 \times 10^{-2}$. Dashed lines indicate the threshold region over the most recent 15 days for day 62. (c) EPLG vs chain length for Heron R1 on day 62 (blue) and day 65 (red). On day 62, there are two 2Q gates at lengths 78 and 79 with poor fidelities due to a TLS interaction involving the qubits in those gates. EPLG recovers by day 65.}
  \label{fig:fig4}
\end{center}

increase confined to the $N$-qubit chain signals a localized fault. In such cases, the per-edge process fidelities directly identify the gates responsible.

We use the optimal chain-finding method described in Section ~\ref{sec:protocol} to find the long $N$-qubit chain mentioned above. As for the shorter $M$-qubit chain, we can pick this chain as a subchain within the $N$-qubit chain that has the lowest EPLG value for length $M$. That is, the subchain with the highest product of process fidelities across 50 connected qubits from all measured fidelities on the 100 qubit-long chain. To detect sudden performance degradation, we define a high outlier threshold as $T=Q_3+ 1.5 \times IQR$ over a rolling 15-day window, where $Q_3$ is the third quartile, $Q_1$ is the first quartile, and $IQR = Q_3 {-} Q_1 $.

We analyze stability on two devices: Heron R1 ($ibm\_torino$, 133 qubits) and Heron R2 ($ibm\_fez$, 156 qubits). Heron R2 differs from R1 in two main respects: it increases the qubit count from 133 to 156, yielding more candidate N- and M-qubit chains, and it incorporates a two-level system (TLS) mitigation mechanism not present in R1 \cite{heronr2}. 

On both devices, EPLG remains stable for the shorter $M$-qubit chain, with no significant fluctuations during the 100-day period (see Fig.\mbox{~\ref{fig:fig4}}). When scaling to longer $N$-qubit chains, Heron R2 maintains stable EPLG values around a median of $4.9 \times 10^{-3}$, whereas Heron R1 exhibits oscillations mainly between $5.4 \times 10^{-3}$ and $ 7.6 \times,10^{-3}$, with some notable values as high as $1.2 \times 10^{-2}$. During this 100-day period, the absolute difference between the lowest and highest EPLG for the long chain in Heron R1 is almost three times larger than for Heron R2, highlighting greater performance variation. 

A notable event occurs on day 62, where Heron R1 shows a sudden EPLG spike to $1.2 \times 10^{-2}$. This anomaly is traced to two specific 2Q gate pairs within the $N$-qubit chain used that day. Further analysis revealed that two qubits involved in those gates were interacting with a TLS during this time frame. TLS defects are known to significantly impact qubit coherence  \mbox{\cite{tls1}} \mbox{\cite{tls2}} \mbox{\cite{tls3}} and gate errors \mbox{\cite{utility}} \mbox{\cite{errormitigation}}, and this effect is directly captured by the layer fidelity benchmark. The fidelity of the associated 2Q gates recovers by day 65 and EPLG goes back to a stable value of $7.5 \times 10^{-3}$ for N = 100 (see Fig.\mbox{~\ref{fig:fig4}} (c)).

Across other Heron R2 devices, we observe similar stability trends to $ibm\_fez$ for the best N- and M-qubit chains, featuring stable EPLG values for both chain lengths. These findings could suggest that the additional TLS mitigation mechanisms in Heron R2 devices \cite{heronr2} contribute to this improved stability, though it's possible that having a larger number of candidate chains from an increased qubit count could also result in more stable chains to choose from.

We next consider the stability of a fixed $N=100$ qubit chain on Heron R1 and R2 devices. To do this, on each device, we start with a chain identified as the optimal chain at time $t = 0$ and track its EPLG variation over the subsequent 100-day period. In this experiment, we did not measure the EPLG of these chains directly; instead, we reconstructed it from historical data containing the process fidelities of all 2Q gates on each device, obtained using the grid method described earlier. The final EPLG values were then computed using Eqs. (2) and (3). This serves as a reliable approximation, as Heron devices exhibit negligible crosstalk \cite{lf}. As expected, Heron R1 shows larger EPLG oscillations than Heron R2. The absolute EPLG difference for Heron R2 is an order of magnitude smaller (see Fig.~\ref{fig:fig5} (a)).

\begin{center} 
    \includegraphics[scale=0.55]{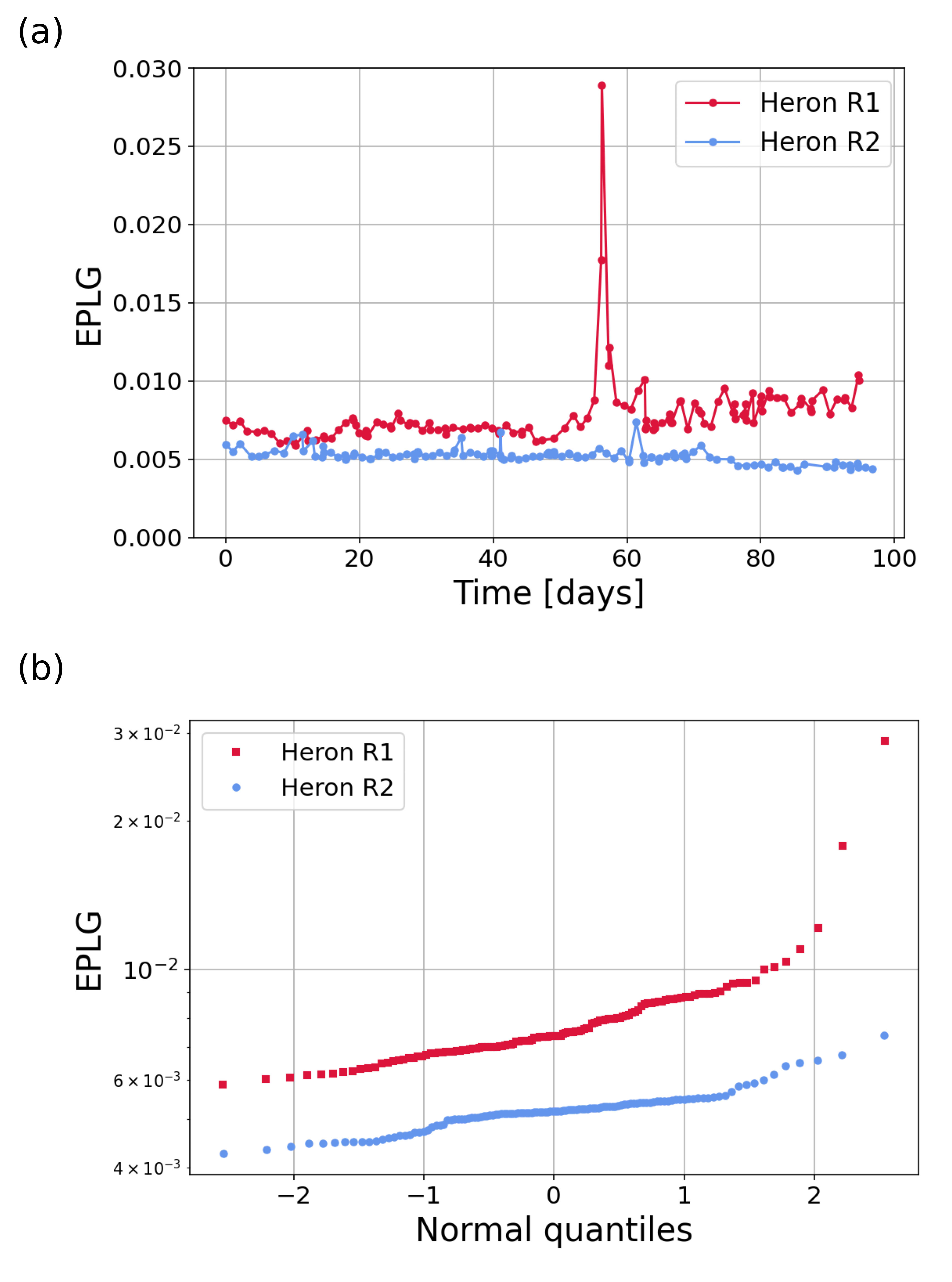}
    \captionsetup{type=figure}
    \captionof{figure}{ EPLG vs time for  (a) Heron R2 ($ibm\_fez$, blue) and (b) Heron R1 ($ibm\_torino$, red) for a fixed chain of length $N=100Q$. On day 62, the EPLG of Heron R1 increases to $2.8 \times 10^{-2}$ due to a strong TLS interaction involving one qubit in the chain. (b) Normal quantile distribution of EPLG values over a 100-day period for these fixed chains.}
\label{fig:fig5}
\end{center}

The same TLS-related event that affected the optimal chain of Heron R1 on day 62 also impacts the fixed chain. One gate in the fixed chain, which was also present in the optimal chain, strongly interacted with a TLS on this day, which is clearly reflected in the EPLG spike in Fig.~\ref{fig:fig5} (a).

Examining the normal quantile distribution of EPLG values over the 100-day period, the most pronounced difference between Heron R1 and R2 devices appears in the third quantile, where the values of Heron R1 deviate from the linear trend observed elsewhere (Fig.~\ref{fig:fig5} (b)).

Overall, while both devices maintain stable performance for shorter chains, Heron R2 demonstrates markedly improved stability when scaling to long $N$-qubit chains. By capturing both gradual trends and isolated anomalies, layer fidelity provides a direct and scalable means of quantifying device stability, offering a valuable tool for assessing the long-term reliability of large-scale quantum processors.

\section{Assessing EPLG variation from direct RB fits}
\label{sec:fits}

The stability analysis demonstrates that EPLG is a reliable metric for tracking performance. However, its accuracy is influenced by the quality of RB decay fits, which depend on parameters such as number of randomizations and Clifford lengths. To ensure reliable EPLG estimates, we investigate how these factors affect fit precision and establish reasonable lower bounds for minimizing uncertainties.

To achieve this, we collect data from 100 randomizations for each gate in a fixed $N=100$ qubit-long chain. The data is obtained from 10 consecutive layer fidelity runs on a fixed chain on Heron R2 ($ibm\_fez$), with each run consisting of 10 randomizations, 200 shots per circuit, and Clifford lengths $C=[1,30,40,60,80,100,150,200,300,400,500,600]$. All circuits are generated using Qiskit and executed via the IBM Quantum cloud interface (see this \href{https://github.com/qiskit-community/qiskit-device-benchmarking/blob/main/notebooks/layer_fidelity_single_chain.ipynb}{notebook} for running layer fidelity on a fixed chain).

To estimate the decay rate $\alpha$, we randomly sample data from $r$ randomizations and fit them to $P(1) = a \times \alpha ^x +b$. Here, we consider all Clifford lengths in $C$. We then take the nominal fit values, along with upper and lower bounds defined by the standard error. Using these three values, we estimate 1Q and 2Q process fidelities according to $F = \frac{1 + (d^2 -1)\alpha}{d^2}$ where $d=2$ for 1Q gates and $d=4$ for 4Q gates. Finally, we compute layer fidelity as described in equation (2),(3) and normalize it to  EPLG according to equation (4).

In a separate analysis, we repeat this process while varying the number of Clifford points by excluding certain lengths in $C$, and fixing the number of randomizations to $r=10$.

Results are shown in Fig.~\ref{fig:fig6}. The top figure presents EPLG as a function of $r$ randomizations for the nominal fit, along with the upper and lower bounds obtained by propagating the standard error in the RB decay fits. The data reveals a clear asymptote at $4.8 \times 10^{-3}$ with uncertainties decreasing as $r$ grows. As expected, EPLG variation is largest for small values of $r$ but its nominal value stabilizes rapidly for $r \geq 6$. The upper and lower bounds start to stabilize around $r \geq 20$, serving as reasonable lower bounds for the number of randomizations on Heron R2 processors. 

The bottom figure further illustrates EPLG as a function of $c$ Cliffords, obtained by excluding specific lengths from $C$ and refitting. The data shows that the nominal  EPLG value stabilizes at around $c = 100$, with a clear asymptote at $5 \times 10^{-3}$. Uncertainties are largest when using fewer Clifford lengths (e.g., $<300$), but they quickly converge with more Cliffords. We note that if the nominal EPLG were to change significantly, a larger number of Clifford lengths would be required to achieve comparable convergence. 

We emphasize that these parameters are not universal and should be interpreted as hardware-specific guidelines rather than prescriptions. In practice, we recommend performing the same convergence analysis shown in Fig.\mbox{~\ref{fig:fig6}} on the target device to determine the minimum number of randomizations and Clifford lengths required for stable EPLG estimation. In practice, convergence can be defined quantitatively by requiring that (i) the relative change in the nominal 

\begin{center}
  \centering
  \includegraphics[scale=0.55]{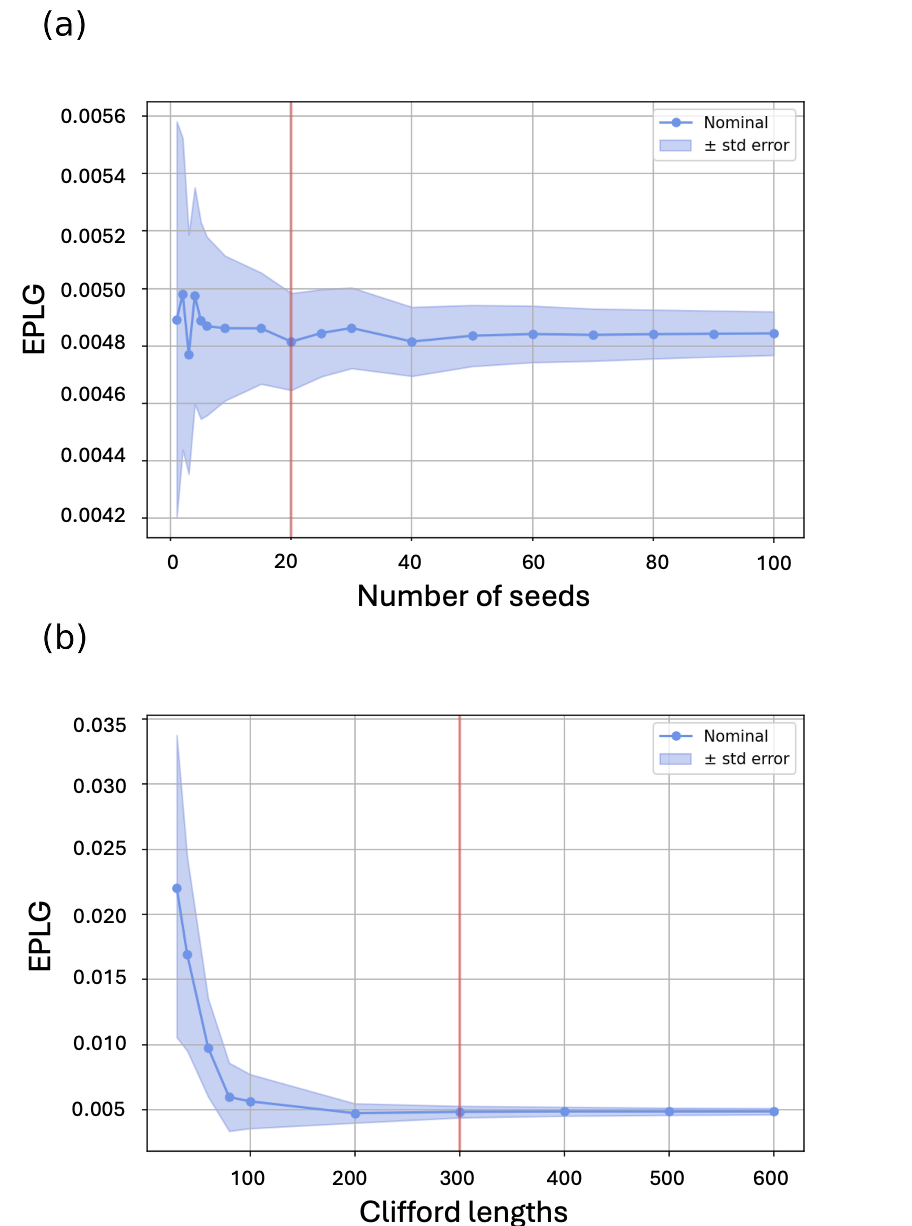}
  \captionsetup{type=figure}
  \captionof{figure}{
  (a) EPLG vs number of randomizations on Heron R2 ($ibm\_fez$) for a chain of length $N=100Q$. Shaded areas denote the standard error, propagated from error fits in RB decay. Nominal EPLG starts stabilizing at around r = 6. Shaded EPLG areas start stabilizing at around r = 20. (b) EPLG vs Clifford lengths on this same backend for a chain of length $N=100Q$. Shaded areas denote the standard error, propagated from error fits in RB decay. Nominal EPLG starts stabilizing at around 100 Cliffords. Shaded EPLG areas start stabilizing at around 300 Cliffords.
  }
  \label{fig:fig6}
\end{center}

EPLG remains below a fixed tolerance when additional randomizations or Clifford lengths are included, and (ii) the propagated uncertainty bounds shrink by less than a comparable threshold over successive increments. Operationally, this corresponds to entering the regime where the RB decay has reached its approximately exponential asymptote and the extracted decay rate $\alpha$ is insensitive to the inclusion of additional long Clifford points. Devices with larger EPLG or greater variability typically require more randomizations (e.g. $>10$) and longer Clifford sets (e.g. $>300$) to achieve comparable convergence. The values reported here therefore serve as practical reference points for superconducting processors of similar scale (e.g., Heron R1,R2 and Eagle R3 processors), but the same procedure can be straightforwardly repeated to optimize these parameters for other architectures.

\section{Analyzing the impact of varying gate lengths}
\label{sec:duration}

While optimizing RB decay fits reduces EPLG uncertainty, other factors also influence layer fidelity results—one key constraint is gate duration. Layer fidelity explicitly enforces barriers at the 2Q gate layer, so all preceding gates must complete before the circuit advances. This mirrors the scheduling used in practical algorithms \cite{utility}, where layered operations cannot begin until the prior layer is finished. Consequently, the effect of lengthened gates is directly reflected in the benchmark, faithfully capturing the timing constraints of real algorithm execution. At the same time, however, enforcing barriers also introduces idle periods for shorter gates in the protocol \cite{lf}, which exposes them to additional decoherence, making it important to examine how this constraint influences the results. On Eagle processors, the fixed coupling architecture imposes stricter constraints on frequency collisions, which often require lengthening certain 2Q gates. In addition, the cross-resonance interaction is highly sensitive to qubit detuning \cite{cr1, cr2, cr3}; when qubits lie in a poor detuning regime, gates must be extended to achieve sufficient entangling interaction.

We compare distributions of 2Q gate errors under three different conditions on two processors that have varying 2Q gate durations: Heron R2 ($ibm\_marrakesh$, median duration 68ns, max duration 80ns) and Eagle R3 ($ibm\_sherbrooke$, median duration 533ns, max duration 881ns). These error distributions are taken from 2Q gates spanning most of the device (see the grid map in Fig.~\ref{fig:fig1} (a) to look at these gates) and are derived from the following cases (see Fig.~\ref{fig:fig7}):

\begin{enumerate}
    \item \textbf{Isolated RB}: A variant of simultaneous RB where 2Q gates are spaced apart by at least two idle qubits. No barriers are applied.
    \item \textbf{Isolated RB (with a delay)}: Same as (1), but each 2Q gate is followed by an additional delay such that the total duration of the layer equals the maximum gate duration in that disjoint layer. This captures length-dependent effects but not crosstalk.
    \item \textbf{Layer Fidelity RB}: Simultaneous RB as described in the layer fidelity protocol. This applies barriers and engages all gates in the current disjoint layer. This captures both crosstalk and length-dependent effects.
\end{enumerate}

Previous work has shown that 2Q gate error distributions are tighter on Herons R2 than on Eagles R3 when considering crosstalk and equal gate durations \cite{lf}. Here, we extend this analysis to include longer gate durations.

We measure all 2Q gate errors on both processors with the methods described above and analyze their distributions using normal quantile plots. In both cases, median errors are lowest for isolated RB, followed by the isolated RB with delay, and then layered RB. 

\begin{center}
  \centering
  \includegraphics[scale=0.8]{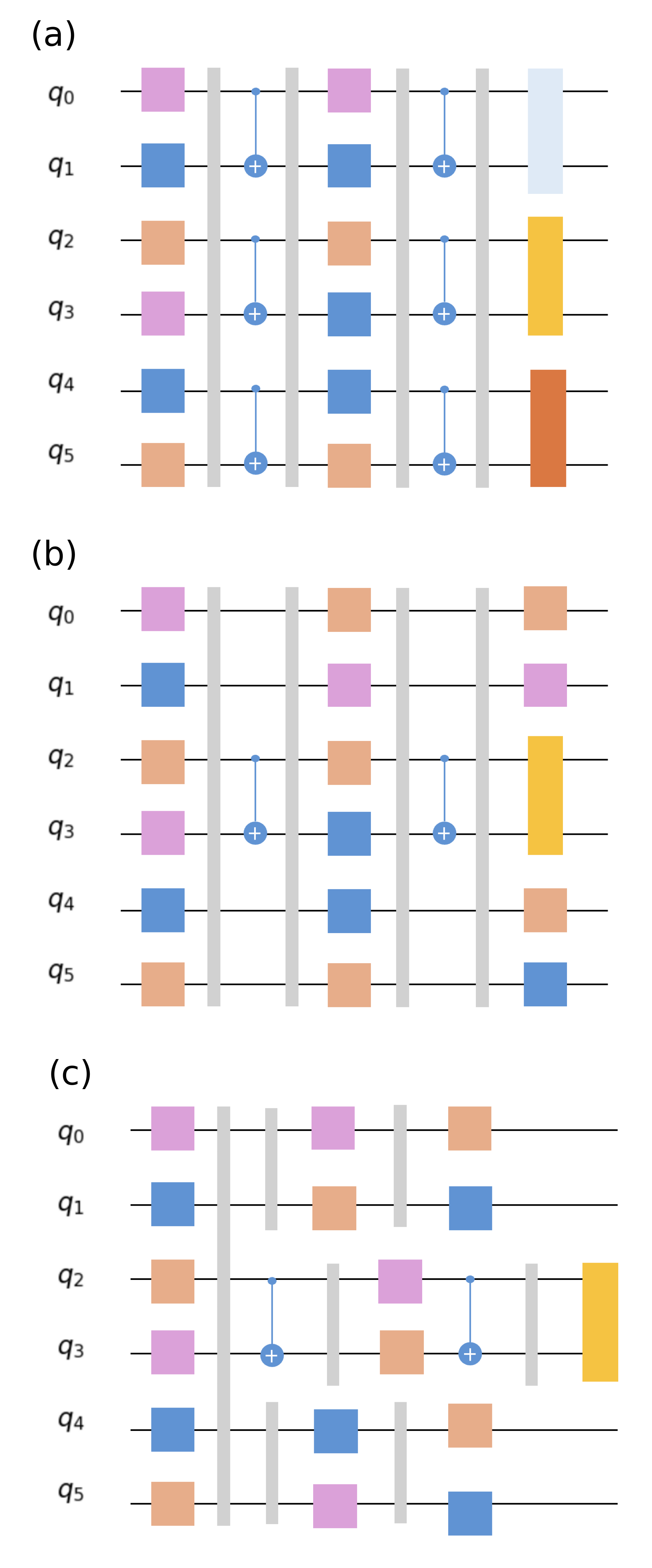}
  \captionsetup{type=figure}
  \captionof{figure}{ The following sections show circuits for a Clifford depth = 2. The span of these circuits is simplified to 6 qubits, but the real circuits span a much larger device region. Measurements are omitted in this diagram, but they are incorporated in the actual experiment. Inverses are included in these circuits and can be found towards the end. (a) Layered RB circuits with timing barriers and simultaneous 2Q gates. (b) isolated RB with a timing delay to simulate the constraint imposed by layer fidelity. (c) isolated RB.
  }
  \label{fig:fig7}
\end{center}

On Heron R2, the differences in median 2Q errors across these distributions are small (see Fig.~\ref{fig:fig8} (a)). On this device, the longest gates is 17\% longer than the median duration. Our data shows that this relative difference in duration has little impact on layered RB outcomes. On Eagle R3, the effect is more pronounced. On this device, the longest gate is 65\% longer than the median duration. We observe that the median isolated RB error with a delay is 2x higher than the isolated RB case with no delay. The layered RB case, which also incorporates crosstalk, notably shows even larger errors (3.9x higher than the median isolated RB case).

\begin{center}
  \centering
  \includegraphics[width=\linewidth]{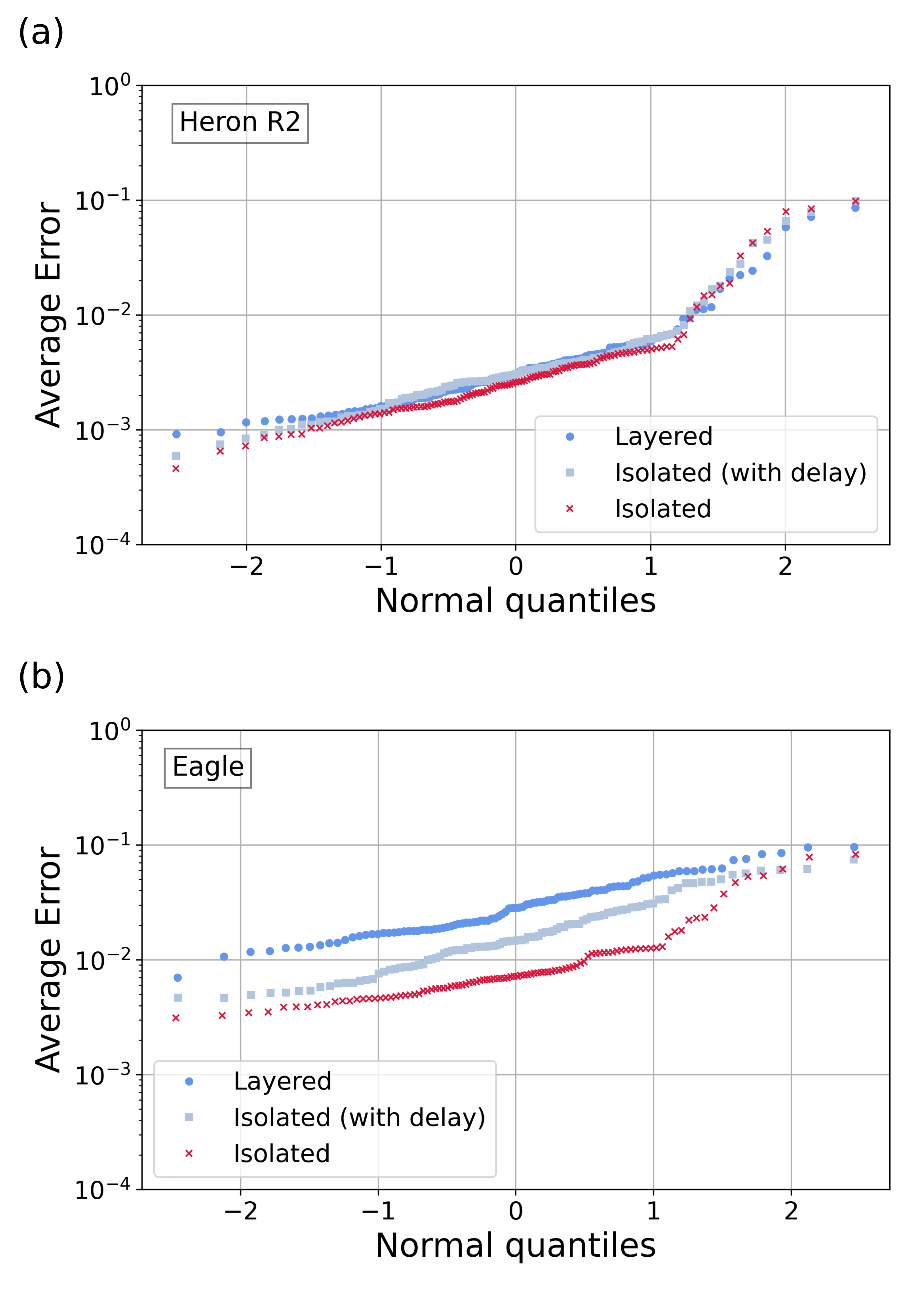}
  \captionsetup{type=figure}
  \captionof{figure}{
   2Q error distributions for Heron R2 ($ibm\_marrakesh$) and (b) Eagle R3 ($ibm\_sherbrooke$). Dark blue line is layered RB, including long gates, light blue line is isolated RB with a delay to impose the same timing constraints as layer fidelity, and red line is isolated RB. Distributions are fairly tight for Heron R1 but differ for Eagle R3. The long gates on this Eagle R3 processor are 65\% longer than the median of the other gates. These distributions isolate the effects of timing barriers and crosstalk on 2Q errors. 
  }
  \label{fig:fig8}
\end{center}

These results highlight that longer 2Q gate durations introduce significant variation on quantum devices, reinforcing the importance of considering these constraints when running quantum workflows on qubit layouts with varying gate lengths.

\section{Conclusion}
\label{sec:conclusion}

In this work, we expand the analysis of the layer fidelity benchmark to enhance its applicability in device performance evaluation. We introduced a protocol for identifying optimal qubit chains of length $N$, demonstrating that our approach yields EPLG values $40 \% - 70 \%$ lower than those from randomly selected chains for $N=100$. While the one-dimensional chain used here was selected for its versatility—it is amenable to many device topologies, easily scaled in size, and well suited for depth-free search—the overall approach can be readily generalized to other layer fidelity topologies. For instance, a square layer (4 disjoint layers) could be adopted for processors with a four-nearest-neighbor square topology, such as the IBM Nighthawk processor. Nonetheless, the chain remains a valuable construct, as it provides a simple yet effective means of capturing device performance trends.

By tracking layer fidelity over time, we also established EPLG as an effective stability monitoring tool, capable of detecting both localized and device-wide performance degradations. Additionally, we refined the accuracy of RB decay fits, determining practical bounds on experimental parameters to minimize fit uncertainties. Because these bounds depend on the underlying hardware and noise characteristics—here evaluated on superconducting processors—they should be re-optimized for other architectures, and practitioners are encouraged to calibrate these parameters for their specific processors.

Finally, we analyzed the impact of gate duration constraints on layer fidelity measurements. We found that longer 2Q gate durations significantly increase layered gate errors on Eagle R3 processors when gate durations were extended by $65\%$. These findings refine the use of the protocol as a quantum benchmarking tool. While our study focuses primarily on superconducting architectures, the underlying methodology is broadly applicable and can be readily adapted to other hardware platforms.

\section*{Acknowledgment}

The authors would like to thank Youngseok Kim, Luke Govia and Seth Merkel for insightful discussions. We also want to thank Oliver Dial for writing a holistic depth-first-search algorithm for finding the best qubit chain on a device. Research was partially supported by the Army Research Office and was accomplished under Grant Number W911NF-21-1-0002. The views and conclusions contained in this document are those of the authors and should not be interpreted as representing the official policies, either expressed or implied, of the Army Research Office or the U.S. Government. The U.S. Government is authorized to reproduce and distribute reprints for Government purposes notwithstanding any copyright notation herein. 

\newpage

\bibliographystyle{unsrt}

\end{multicols}
\end{document}